\input harvmac.tex

\noblackbox
\lref\JCardy{J. Cardy, Nucl. Phys. B324 (1989) 581.}
\lref\GZ{S. Ghoshal, A. B. Zamolodchikov, Int. J. Mod. Phys.
A9, (1994) 3841.}
\lref\AK{C.Ahn, W.M. Koo, Nucl. Phys. {\bf B}468, (1996) 461.}
\lref\P{P. Pearce, Phys. A30 (1997) 2353-2366}
\lref\MISC{A. Leclair, Phys. Lett. 230B (1989) 103;N. Yu
Reshetikhin,
F. Smirnov, Comm. Math. Phys. 131 (1990) 157.}
\lref\PREV{P. Fendley, H. Saleur, J. Phys. A27 (1994) L789.}
\lref\LMSS{A. Leclair, G. Mussardo, H. Saleur, S. Skorik,
Nucl. Phys. {\bf B}453, (1995) 581.}
\lref\ZZ{A.B. Zamolodchikov, Al. B. Zamolodchikov, Nucl. Phys. B358
(1991) 524.}
\lref\FS{P. Fendley, H. Saleur, in Proceedings of the 1993  Trieste
Summer School, World Scientific}
\lref\BS{H. Saleur, M. Bauer, Nucl. phys. B 320 (1989) 591.}
\lref\JC{J. Cardy, Nucl. Phys. B324 (1989) 591.}
\lref\F{P. Fendley, Phys. Rev. Lett. 71, (1993) 2485.}
\lref\BLZ{V.V. Bazhanov, S.L. Lukyanov, A.B. Zamolodchikov,
Comm. Math. Phys.  177, (1996) 381.}
\lref\FLS{P. Fendley, F. Lesage, H. Saleur, J. Stat. Phys. 85 (1996)
211.}
\lref\Dorey{P. Dorey,  A. Pocklington,  R. Tateo , G. Watts,
hep-th/9712197}
\lref\AL{I. Affleck, A. W. W. Ludwig, Phys. Rev. Lett. 67 (1991)
161.}

\Title{\vbox{\baselineskip12pt
\hbox{USC-98-003}}}
{\vbox{\centerline{Boundary flows}
 \vskip 4pt
\centerline{in minimal models.}}}

\centerline{F. Lesage, H. Saleur\foot
{Packard Fellow} and P. Simonetti}
\bigskip\centerline{Department of Physics}
\centerline{University of Southern California}
\centerline{Los Angeles, CA 90089-0484}

\vskip .3in
We discuss in this paper the behaviour of minimal models of
conformal
theory perturbed by the operator $\Phi_{13}$ at the
boundary.  Using the RSOS restriction of the sine-Gordon model,
adapted to the boundary problem, a series of boundary flows between
different set of conformally invariant boundary conditions are
described. Generalizing the ``staircase'' phenomenon
discovered by  Al. Zamolodchikov,
we find that an analytic continuation of the boundary sinh-Gordon
model
provides a  flow interpolating not only  between all minimal models
in the bulk, but also between  their possible conformal
 boundary conditions. In the particular case where the bulk
sinh-Gordon
coupling is turned to zero, we obtain a boundary roaming trajectory
in the
$c=1$ theory, that interpolates between all the possible spin $s$
Kondo models.

\Date{1/98}

\newsec{Introduction}

The problem of minimal models perturbed by the $\Phi_{13}$ operator,
both in the bulk and at the boundary, has been considered
by various groups \AK,\P. Nevertheless,
various aspects are  far from clear: for instance, the physical
meaning of the solutions of the boundary Yang Baxter equation
that have been found, or the relation with the boundary sine-Gordon
model via quantum group truncation, if any.

This is not only of academic
interest. Minimal models arise naturally in quantum impurity
problems, when one considers, for instance,
situations involving several channels through a conformal embedding
approach. When the bulk is massless and the boundary still
perturbed by $\Phi_{13}$, one
gets a boundary flow in the minimal model, ie an interpolation
between two different types of boundary conditions (boundary fixed
points).
What sort of boundary conditions
can be related in that way in the  corresponding
quantum impurity problem, and what is
the boundary free energy along the flow,
are questions of crucial importance (for recent progress in that
direction
using a different approach, see
\Dorey.)

Consider the most general problem of  bulk and
boundary perturbation of minimal models.
By following the standard arguments
\GZ, one finds that the theory
with hamiltonian
\eqn\pertmin{H=H_{CFT}+g\int_{-\infty}^0 dx \Phi_{13}\bar{\Phi}_{13}
+g_B \Phi_{13}(0),
}
is integrable for any choice of the coupling constants $g,g_B$.
Here,
$H_{CFT}$ is the fixed point hamiltonian of a minimal model
with central charge $c=1-{6\over m(m+1)}$,
and $\Phi_{13}$ is the field with dimension $h_{13}={m-1\over m+1}$.

In the case of a bulk theory, and for $g>0$ (so the bulk is massive)
it is well understood how the
problem can be considered as a truncation of the sine-Gordon model
\MISC\ at coupling ${\beta^2\over 8\pi}={m\over m+1}$.
The simplest way to think of this is the scattering theory, where
the
truncation amounts to going from the soliton S-matrix
(a vertex type solution of the
Yang Baxter equation) to  the kink  S-matrix (an interacting round a
face type solution)
via quantum 6j calculus based on the algebra $sl_q(2)$,
$q=\exp\left(i\pi
{m-1\over m}\right)$.
In this process,
it is of crucial importance that the soliton S-matrix  has a quantum
group symmetry, in particular that it conserves the $U(1)$  charge.

It therefore appears  very natural to try to describe the theory
\pertmin\ with both
a bulk and a boundary perturbation as a truncation of the boundary
sine-Gordon model
with hamiltonian
\eqn\pertsg{H=H_{FB}+ G\int_{-\infty}^0 dx\cos\beta\phi+G_B
 \cos{\beta\over 2}(\phi-\phi_0)(0).}
Here $H_{FB}$ is the free boson hamiltonian. The coupling constants
in \pertsg\ are related with the ones in \pertmin\ by $G\propto
\sqrt{g}$.

One quickly meets a major difficulty however: while the bulk
S-matrix
does
conserve the
$U(1)$ charge, the boundary R matrix, as worked out in \GZ, does not
in
general\foot{The only limiting cases where there is a conserved
$U(1)$
charge are the case of   Dirichlet boundary conditions,
 $G_B=\infty$, and the case of Neumann boundary conditions
when the bulk is massless, $G=G_B=0$. These correspond to boundary
fixed points
 in minimal
models, and do not describe the {\sl flows} one is looking for.}.
 It is thus impossible to implement the quantum group truncation to
obtain
a boundary
scattering for the kinks, and the question occurs, of whether the
quantum group
approach is any good to describe  boundary perturbation of minimal
models.

The purpose of this note
is to demonstrate that the problem \pertmin\ {\sl can}  be
approached
using
sine-Gordon type models, but that to do so, one has to include
boundary degrees of freedom
of the Kondo type (see \F\
for a discussion of the Kondo problem in the integrable
field theory  language). The idea is not so new, and was proposed
independently
and  in a slightly
different context by N. Warner in the case of supersymmetric models
\ref\npw{N. Warner, Nucl. Phys. B450 (1993) 413.}. Some of the
following results
have appeared in very succint form in previous papers of some
of the authors \PREV, \ref\alllll{A. Leclair, G. mussardo, H. Saleur,
S. Skorik, Nucl. Phys. B453 (1995) 581.}.

\newsec{Massless bulk}

We start by  considering the case where the
bulk is massless, ie $g=G=0$. As is well known, this can be
described
using a massless
scattering theory \ZZ,\FS. In the sine-Gordon case, one has massless
solitons and antisolitons
which are either R or L moving, with corresponding dispersion
relation
$e=\pm p=\mu e^\beta$, $\beta$ the rapidity, $\mu$ an arbitrary
energy scale,
 and scattering given by
$S_{RR}=S_{LL}=S_{SG}$,
the usual SG S -matrix, while the LR scattering is a simple constant
phase. The
truncated version of this is immediate: one has R or L moving kinks,
with
$S_{RR}=S_{LL}=S_{min}$. Now, as far as the boundary goes,
we claim first that action \pertmin\ does not make sense without
specifying the
UV boundary conditions. We will restrict to conformal invariant
boundary conditions (integrable flows only interpolate between
such fixed points).
Then, a  set of conformal boundary conditions was obtained in
\BS,\JC. The simplest way
 to characterize
them is to use the microscopic description of minimal models, where
 ``height'' variables take values on a $A_m$ Dynkin diagram,
$a=1,\ldots,m$.
The first type of boundary condition is obtained simply by fixing
the heights to a constant; it corresponds, in the continuum limit,
to
a boundary state
$|\tilde{h}_{1a}>$ in the  notations of \JC. This state
differs from
$|\tilde{h}_{1, m+1-a}>$ by the phase of some of the coefficients
only, as a result of
the $Z_2$ symmetry of the $A_m$ diagram. In the following, we will
restrict to the case
$a\leq {m+1\over 2}$.
 The second type of boundary conditions
is obtained by fixing the boundary heights to a value $b$ and in
addition,
their neighbours one layer in from the boundary to a value $c$. For
$b,c\leq {m+1\over 2}$, setting $d=\inf(b,c)$,
the corresponding boundary state is $|\tilde{h}_{d1}>$ (again, this
state
differs from $|\tilde{h}_{m-d,1}>$ by the phase of some coefficients
only). For
$(b,c)=\left({m\over 2},{m\over 2}+1\right)$ and $m$ even, the state
is $|\tilde{h}_{m/2,1}>$ .  In this paper,
we will solve the problem of perturbing minimal models with the
first
type of boundary conditions.

Our first claim is that the perturbation of the minimal model with
UV
boundary
conditions of the type $|\tilde{h}_{11}>$ is, indeed, described by
the only possible quantum group
reduction of the boundary sine-Gordon model, that is, is trivial. A
simple
illustration of this is provided by the case $m=3$, the Ising model.
In that case, our
boundary condition corresponds to fixed spins, while the
boundary  operator $\Phi_{13}$ corresponds to the boundary magnetic
field:
clearly, perturbing
fixed spins by  a boundary field is a trivial operation.

Consider now the UV boundary condition $|\tilde{h}_{1a}>$
($a\leq {m+1\over 2}$), and set
 $a=1+j$. We conjecture
that the problem is described by the quantum group truncation of the
following hamiltonian
\eqn\kond{H=H_{FB}+G_B \left(S^+ e^{-i\beta_{SG}\phi(0)/2}+S^-
e^{i\beta_{SG}\phi(0)/2}\right),
}
where the opertors $S^\pm$ are in the spin $j/2$ representation of
the quantum
group $sl_{q'}(2)$, $q'=\exp\left(i\pi{m\over m+1}\right)$, and
${\beta_{SG}^2\over 8\pi}={m\over m+1}$. For $G_B$
real, ie $g_B>0$, this problem
is nothing but the anisotropic spin $j/2$ Kondo problem, which is
well known
to be integrable \ref\kondo{P. Wiegmann, J. Phys. C. 14 (1981)
1463.}\FLS,\BLZ.
 The R matrix is given by the Yang Baxter solution
based on $sl_q(2)$ for scattering a  spin $1/2$ at rapidity $\beta$
through a spin $(j-1)/2$ at some fixed rapidity $\beta_B$
(multiplied
by the appropriate factors to ensure crossing and unitarity); the
value of
$\beta_B$ is related with the coupling constant,
 $\mu e^{\beta_B}\propto g_B^{m+1\over 2}$.
 Note the renormalization
 of the quantum group
parameter and of the spin. The latter occurs because the R matrix
description
is really an IR one, and in the IR, the spin $j/2$ has been
partially
screened by the bulk degrees of freedom, so only the remainder
$(j-1)/2$
appears. The case $j/2=1/2$ deserves special mention. In that case,
there is no
left over spin, and the R matrix is a simple CDD factor,
$R=-i\tanh\left(
{\beta-\beta_B\over 2}-{i\pi\over 4}\right)$.

The quantum group truncation of this R matrix is straightforward: it
becomes
a solution of Yang Baxter for scattering a kink with spin $1/2$
adjacency
rules through a kink with spin $(j-1)/2$ adjacency rules. In the
case
$j/2=1/2$, we
 still
get a CDD factor. To summarize
\eqn\summ{\eqalign{R=&Id,\ a=1\cr
R=&-i\tanh\left(
{\beta-\beta_b\over 2}-{i\pi\over 4}\right),\ a=2\cr
R=& R^{1/2,(a-2)/2}(\beta-\beta_b),\ 3\leq a\leq {m+1\over 2}.\cr}}

The boundary free energy can easily be read off from the
thermodynamic Bethe
ansatz \ref\tbapapers{Al. B. Zamolodchikov, Nucl. Phys. B342 (1990)
695.}.
 As in the bulk case, we simply truncate the TBA for the
bosonic theory -
here  the anisotropic  Kondo model.
Introduce
the $A_{m-2}$ Dynkin diagram with nodes labelled by an integer, and
incidence
matrix $N_{pq}$ ie $N_{pq}=1$ if the nodes $p$ and $q$ are
connected,
zero
otherwise. Introduce the pseudo energies $\epsilon_p$ solutions of
the system
\eqn\tba{\epsilon_p(\beta)=\mu e^\beta\delta_{p1}-T\sum_k N_{pq}
\int {d\beta'\over 2\pi}{1\over\cosh(\beta-\beta')}
\ln\left(1+e^{-\epsilon_q(\beta')}\right).}
Then one has, for the perturbation of the conformal boundary
condition
 $|\tilde{h}_{1a}>$,
with $2\leq a\leq {m+1\over 2}$ and $g_B>0$,
\eqn\bdrf{f^{(a)}_{bdr}=-T\int{d\beta\over 2\pi}
{\ln\left(1+e^{-\epsilon_{a-1}(\beta)}\right)\over
\cosh(\beta-\beta_B)}.}
Of particular interest are the boundary entropies \AL\ in the UV and
the
IR, related
to the degeneracy factors by $s=\ln g$. From \bdrf, one has
$$
{g_{UV}\over g_{IR}}={1+e^{-\epsilon_{a-1}(-\infty)}\over
1+e^{-\epsilon_{a-1}(\infty)}}
$$
In the limit $\beta\to\pm\infty$, the $\epsilon$'s go to constants.
Setting
$x=e^{-\epsilon(-\infty)}$, we have to solve the system
$$
x_p=(1+x_{p-1})^{1/2}(1+x_{p+1})^{1/2}
$$
where $x_0=0$  and $x_{m-1}=0$. The solution is well known
$$
x_p=\left({\sin {\pi(p+1)\over m+1}\over \sin {\pi\over
m+1}}\right)^2-1.
$$
Similarly, for $\beta=\infty$, we have to solve  a similar  system
but
with one less
node, whose  solution is thus
$$
y_p= \left({\sin {\pi p\over m}\over \sin {\pi\over m}}\right)^2-1
$$
It follows that
\eqn\degfct{{g_{UV}\over g_{IR}}=
{\sin {\pi a\over m+1}\over \sin {\pi\over m+1}}
{\sin {\pi\over m}\over \sin {\pi (a-1)\over m}}.}
On the other hand, recall the formula for the degeneracy factors of
the boundary
states
$$
g_{1a}=\left[{2\over m(m+1)}\right]^{1/4}(-1)^{a+1}
\left(2{\sin {\pi \over m}\over \sin {\pi\over m+1}}\right)^{1/2}
\sin{\pi a\over m+1}
$$
and
$$
g_{d1}=\left[{2\over m(m+1)}\right]^{1/4}(-1)^{d+1}
\left(2{\sin {\pi \over m+1}\over \sin {\pi\over m}}\right)^{1/2}
\sin{\pi d\over m}
$$
We thus see that $g_{UV}/g_{IR}=g_{1a}/g_{a-1,1}$. From the
microscopic interpretation  given above, this is a flow from
fixed boundary conditions on one row ($a$) to fixed boundary
conditions
($a,a-1$) on two neighbouring rows. Alternatively, by considering
the
row one
layer in,
we can think of it as a flow from a boundary condition where two
heights $a\pm 1$
are allowed, to one where heights are fixed to the value $a-1$. It
is
easy to check  that the dimension of the UV perturbing operator
is indeed
$h_{13}$, since  the diagram has the same skeleton
as for the bulk perturbation. One also checks that the IR fixed
point
is approached along the direction $\Phi_{31}$.

When $a>  {m+1\over 2}$, of course, one  can  still use the previous
results
by exploiting the symmetry of the lattice
problem under $a\to m+1-a$. In microscopic terms,
the flow is now from one height $a$ fixed  to
two neighbouring heights fixed: $a,a+1$.

Nevertheless, the TBA diagram is not  $Z_2$ symmetric because the
source term is only on the left. Therefore, if we use the same
$R$ matrices as before with $a> {m+1\over 2}$, we  will describe
another flow.
By comparing with the boundary entropies,
it is a flow from  $|\tilde{h}_{1a}>$
($\equiv|\tilde{h}_{1,m+1-a}>$)
to $|\tilde{h}_{a-1,1}>$ ($\equiv|\tilde{h}_{m+1-a,1}>$).
In microscopic terms, setting $a'=m+1-a$, $a'\leq {m+1\over 2}$,
we have  a flow from one  height  fixed to $a'$ to heights fixed
$a',a'+1$

It is  reasonable to expect that this flow corresponds to the
perturbation with
$g_B<0$, and this can be proven easily by considering the functional
relations
approach to the problem \BLZ,\FLS. Setting $f_{bdr}^{(a)}=-T\ln
Z_{j}$, $a=1+j$,
the $Z_{j}$ can be shown to obey fusion relations of the type
\eqn\fusion{Z_1(G_B)Z_{j}(G_B)=Z_{j-1}(q'^{1/2}G_B)
+Z_{j+1}(q'^{-1/2}G_B).}
For minimal models, these relations close, and one has
\eqn\closure{Z_j(G_B)=Z_{m-1-j}(iG_B).}
If we expand the boundary free energy in powers of
the coupling constant $g_B\propto G_B^2$ for either case, we get
indeed identical
 expansions
but with odd terms having the sign switched. To summarize,
we have, for $ a\leq {m+1\over 2}$,
\eqn\bdrfi{\eqalign{f^{(a)}_{bdr}(g_B>0,g=0)=&-T\int{d\beta\over
2\pi}
{\ln\left(1+e^{-\epsilon_{a-1}(\beta)}\right)\over
\cosh(\beta-\beta_B)},\hbox{ flow }a\to (a,a-1)\cr
f^{(a)}_{bdr}(g_B<0,g=0)=&-T\int{d\beta\over 2\pi}
{\ln\left(1+e^{-\epsilon_{m-a}(\beta)}\right)\over
\cosh(\beta-\beta_B)},\hbox{ flow }a\to (a,a+1)\cr
}}
where the $\epsilon$'s solve the equations \tba. As in the bulk
case,
we expect the correspondence between the parameter $\beta_B$ and
$g_B$ to be the same in both regimes up to a sign,
 $\mu e^{\beta_B}\propto |g_B|^{m+1\over 2}$.

\newsec{Massless flow in the bulk}

It is interesting to discuss the related problem where one has
a massless flow both in the bulk and at the boundary. In that case,
the bulk scattering is well known, and involves diffusion of left
and right particles in a non trivial way \ref\zamomassless{Al. B.
Zamolodchikov, Nucl. Phys. B366 (1991) 122.}. We conjecture that the
TBA
is given by the following.  Introduce the pseudo energies
$\epsilon_p$
solutions of the system
\eqn\tba{\epsilon_p(\beta)=\tilde{\mu}
e^\beta\delta_{p1}+\tilde{\mu}
e^{-\beta}\delta_{p,m-2}
-T\sum_k N_{pq}
\int {d\beta'\over 2\pi}{1\over\cosh(\beta-\beta')}
\ln\left(1+e^{-\epsilon_q(\beta')}\right).}
Then one has, for the perturbation of the conformal boundary
condition
$|\tilde{h}_{1a}\rangle$
with $1<a\leq {m+1\over 2}$
\eqn\bdrfii{f^{(a)}_{bdr}(g_B>0,g<0)=-{T\over 2}\int{d\beta\over
2\pi}
\left[{\ln\left(1+e^{-\epsilon_{a-1}(\beta)}\right)\over
\cosh(\beta-\beta_B)}+
{\ln\left(1+e^{-\epsilon_{m-a}(\beta)}\right)\over
\cosh(\beta+\beta_B)}\right].}
The mass scale $\tilde{\mu}$ now has a direct physical meaning. The
other
 scale in the problem is the boundary scale $T_B=\tilde{\mu}
e^{\beta_B}$.

Various checks are available for this conjecture.
First, consider the case where $\tilde{\mu}\to 0$, so there actually
is
no flow in the bulk, or, more precisely, the whole boundary flow
takes place at energies in the deep UV of the bulk flow. Then, we
have to
send $\beta_B\to\infty$ to keep $T_B$ finite. In doing so, the TBA
factorizes
in two parts, corresponding respectively to the limits $\beta\to
\pm\infty$.
In the first case, setting $T_B=\mu e^{\theta_B}$ and $\tilde{\mu}
e^\beta=\mu e^{\theta}$,
the mass term on the $m-2^{th}$ node of the TBA disappears,
and we get for the boundary free energy a contribution exactly
equal to half the one in \bdrf, with $\beta_B\equiv \theta_B$. The
left moving part
of the TBA gives the same result, so the two add up to
what we had previously (\bdrf), as desired.

Another interesting case is  what happens at fixed  $\beta_B$
when
$\tilde{\mu}$ is varied. As $\tilde{\mu}\to 0$, the boundary is in
the UV limit for
the $A_m$ model,
while as $\tilde{\mu}\to\infty$, it should be in the IR of the
$A_{m-1}$
model. From our formulas,
when $\tilde{\mu}\to 0$, the two source terms in the TBA disappear,
so we get, using results of the previous section
\eqn\mixedi{f_{bdr}=-T\ln\left({\sin{\pi a\over m+1}\over
\sin{\pi\over
m+1}}\right).}
On the other hand, as $\tilde{\mu}\to\infty$, the two nodes  of the
TBA
carrying source terms
 disappear,
so
\eqn\mixedii
{f_{bdr}^{(a)}=-T\ln\left({\sin{\pi (a-1)\over m-1}\over
\sin{\pi\over
m-1}}\right).}
The difference of \mixedi\ and \mixedii\ corresponds to a flow
from $|\tilde{h}_{1,a}\rangle$ in the $A_m$ model to
$|\tilde{h}_{a-1,1}\rangle$
 in the $A_{m-1}$
model, as expected.

Finally, we can also consider the case where $\tilde{\mu}$ is very
large and
 $\beta_B\to-\infty$,
so $T_B$ is finite. Setting $T_B\equiv \mu e^{\theta_B}$, by varying
$T_B$
we should describe the boundary flow in the $A_{m-1}$ model. Indeed,
in that limit, for the first integral of \bdrfii, what contributes
is
the limit $\beta\to-\infty$. Setting $\tilde{\mu} e^{\beta}\equiv
\mu e^{\theta}$
then,
we get the same expression as \bdrf, while for the TBA the
$m-2^{th}$
node has disappeared. A similar thing occurs for the second
integral,
so we  get indeed the same thing as \bdrf, but with one less node
for
the
TBA, as expected.

Of course, one can also check the limit where one is always in the
UV
of the boundary flow, or where one is always in the IR of this flow.

The conjecture has an immediate extension to the case when $g_B<0$:
\eqn\bdrfiii{f^{(a)}_{bdr}(g_B<0,g<0)=-{T\over 2}\int{d\beta\over
2\pi}
\left[{\ln\left(1+e^{-\epsilon_{a-1}(\beta)}\right)\over
\cosh(\beta+\beta_B)}+
{\ln\left(1+e^{-\epsilon_{m-a}(\beta)}\right)\over
\cosh(\beta-\beta_B)}\right].}
We leave it to the reader, to check that the various limiting cases
are correctly reproduced.

\newsec{Boundary roaming trajectories.}

In \ref\zamoroam{Al.B.  Zamolodchikov, ``Resonance factorized
scattering
and roaming trajectories, unpublished preprint.}, Al. Zamolodchikov
introduced
a scattering theory with a single bosonic particle (and  no bound
state)
which appears to interpolate (``roam'')
 between successive minimal models, exhibiting plateaux where the
running
central charge takes values $c=1-{6\over m(m+1)}$
as the RG scale is
varied.

This scattering theory can be formally considered as  an analytic
continuation to complex dimensions of the sinh-Gordon model
\eqn\sinhg{
H=H_{FB}+\nu \int_{-\infty}^\infty dx \cosh\beta\phi.
}
Defining the coupling constant
$$
B={1\over 2\pi}{\beta^2\over  1+{\beta^2\over 4\pi}}
$$
the  continuation $B\rightarrow 1\pm
{2i\over \pi}\theta_0$ in the sinh-Gordon S matrix coincides indeed
with the scattering theory in \zamoroam\ . Under this analytic
continuation,
each choice of sign leads to a different complex dimension
$\Delta_\pm$
for  the perturbation in \sinhg: it was argued in \zamoroam\ that the
action corresponding to the continued scattering theory
involves a real  combination of both choices.

Though the results of \zamoroam\ are still to a large extent
mysterious,
it is interesting to see whether they extend to the boundary case.
We thus consider the sinh-Gordon model \sinhg\ in the half space
only,
and add a boundary term
\eqn\bact{
H_B=\lambda  \cosh{\beta\over 2}(\phi (0)-\phi_0).
}

The  boundary reflection matrix is known for real $\beta$
\ref\ghoshal{S. Ghoshal, Int. J. Mod. Phys. A9 (1994) 4801.};
defining
\eqn\parran{
\xi (a)={\sinh({\theta\over 2}+i{\pi a\over 4})\over
\sinh({\theta\over 2}-i{\pi a\over 4}),}
}
the reflection matrix is given by
\eqn\reflecmat{
R={\xi (1)\xi (2-B/2)\xi (1+B/2)\over
\xi (1-E)\xi (1+E)\xi (1-F)\xi (1+F)}
}
with $E$ and $F$ related to the boundary parameters
$\lambda$ and $\phi_0$ in a complicated way.

We now wish to formally extend the theory to complex values of
$\beta$,
by continuing   the coupling $B$ in the bulk and boundary scattering
matrices; for the reflection matrix to remain a pure phase, one
needs to adjust the scales $E,F$ appropriately
in the complex plane.

To compute the full boundary free energy, all terms of order $1/L$
($L$ the size
of the system) have to be carefully taken into account. We are
mostly
interested in
changes of boundary entropy on the bulk plateaux, and for this,
simpler formulas  are sufficient.
Introducing the bulk TBA
\eqn\epsilnn{
-mR\cosh\theta+\epsilon+{1\over 2\pi}\phi *
\log(1+e^{-\epsilon})=0,
}
where the kernel is
\eqn\kernbulk{
\phi=-i\partial_\theta \log S={1\over \cosh(\theta+\theta_0)}+
{1\over \cosh(\theta-\theta_0)},
}
one has
\eqn\entropy{
s_B(E,F)={1\over 4\pi}\int d\theta \
\kappa(\theta) \log(1+e^{-\epsilon(\theta)}),
}
where $\kappa=-i\partial_\theta\log R$.  The boundary flow at a
fixed
value of
$mR$  is
completely described by the parameters $E,F$ .
For the rest of this work, we will further restrict to the
case $E=0$, corresponding to $\phi_0=0$ in the
sinh-Gordon action, and the previous $\phi_{13}$ considerations. In
that
case one has
\eqn\bdmatsimp{
\kappa(\theta)={2\cosh{\pi F\over 2} \cosh\theta
\over (\cosh^2\theta-\sinh^2{\pi F\over 2})}
}

To fix the ideas, let us start by reproducing one of the results
of \zamoroam:  the function $c(mR=r)$
is plotted for $\theta_0=50$ in Fig. 1.
As can be seen, this function
interpolates between the values for successive
minimal models, starting from the
Ising model  with $c={1\over 2}$ ($m=3$).

Let us now position ourselves on one of these plateaus by
choosing $r$ appropriately. We can then study what happens to
the boundary entropy as we change $F$, which describes
the scale of the boundary coupling constant $\lambda$.  
The results are shown  in  Fig. 2  for the first three plateaus.

A detailed study of the numerical values of the boundary entropies
reveals that the boundary trajectory interpolates between all
allowed flows in the minimal models discussed in section 2,
though (roughly) half of these flows are ``inverted''.
Let us discuss this further. Consider first the case $c=1/2$:
the flow observed in Fig. 2 has a boundary entropy
of $\log(2)/2$ in the UV, and zero in the IR,
corresponding to the flow from (in microscopic variables)
$2$ to $2,1$ in the $A_3$ model (free to fixed boundary conditions
in
the usual Ising model).
The next plateau shows more structure: the boundary
entropy starts with a plateau at $\log(2)/2$
and then {\sl rises}  to another plateau
with value $\log(2\cos{\pi\over 5})$:
the entropy difference is the opposite of the  one for the flow from
$2$
to $2,3$
in the $A_4$ model: in the physical flow the boundary
entropy would decrease, while here it increases.
The running entropy  then decreases to
zero, and the new entropy difference is the one for the
flow from $2$ to $2,1$ in the $A_4$ model.
The next curve for $c=4/5$ is getting even more intricate
but each difference between the plateaus corresponds to one
of the flows described in section 2.
{}From left to right, the difference of successive entropies
correspond to flows from $3$ to $3,2$, the opposite of the flow
from $2$ to $2,3$ and
finally the flow from $2$ to $2,1$ in the $A_5$ model.
The same pattern
have been verified for the next two plateaus of $c$,
corresponding to $m=6,7$.

These observations are somewhat satisfactory, bringing additional
evidence that the roaming trajectory truly interpolates between
minimal
models. It is quite mysterious that the entropy actually
increases along some of the flows, though this is not forbidden by
the ``g-theorem''\ref\AL{I. Affleck, A. Ludwig, Phys. Rev.
Lett. 67 (1991) 161.}, since the theory we are dealing with
involves complex dimension.
The pattern is that, every time the sinh-Gordon model tries
to ``mimic'' a flow
of
the type
$a\to a,a-1$ ($a\leq {m+1\over 2}$) in the $A_m$ model,
it does it right, but every time it tries
to mimic the flow $a\to a,a+1$, it does it in reverse, producing
a flow $a,a+1\to a$ instead.

It is especially interesting to consider the case  $m\to\infty$, ie
when the
bulk coupling vanishes. In that case, one expects a roaming within
the $c=1$ possible boundary conditions. Though this is difficult to
study numerically,
it is easy to find out what happens by taking the $m\to \infty$
limit. For
fixed $a$, as $m\to\infty$, the ratio of $g$ factors for the
inverted
flow
$a,a+1\to a$ goes to unity, while the ratio for the normal
flow $a\to a,a-1$ goes to ${a\over a-1}$. The boundary entropy
therefore interpolates between the
logarithms of successive integers.
Integer boundary degeneracies are well known
in the $c=1$ theory: they correspond to Kondo models with spin $j$,
and $g=2j+1$. It is therefore tempting to conjecture that we have
here a trajectory interpolating between successive spin $j$ Kondo
models,
all  the way from $j=\infty$ in the deep UV ($F\to 0$) to $j=0$
($F\to\infty$)
in the IR

\vskip4pt
\noindent{\bf Acknowledgments}: we thank W. M. Koo and A. Leclair
for
useful discussions.

\listrefs

{\centerline \rm Figures Captions.}
\vskip 0.5cm

{\rm Fig. 1:} Roaming trajectory for $\theta_0=50.$ 
\vskip 0.4 cm

{\rm Fig. 2:} Boundary flows for the three first plateaus of $c(mR)$.

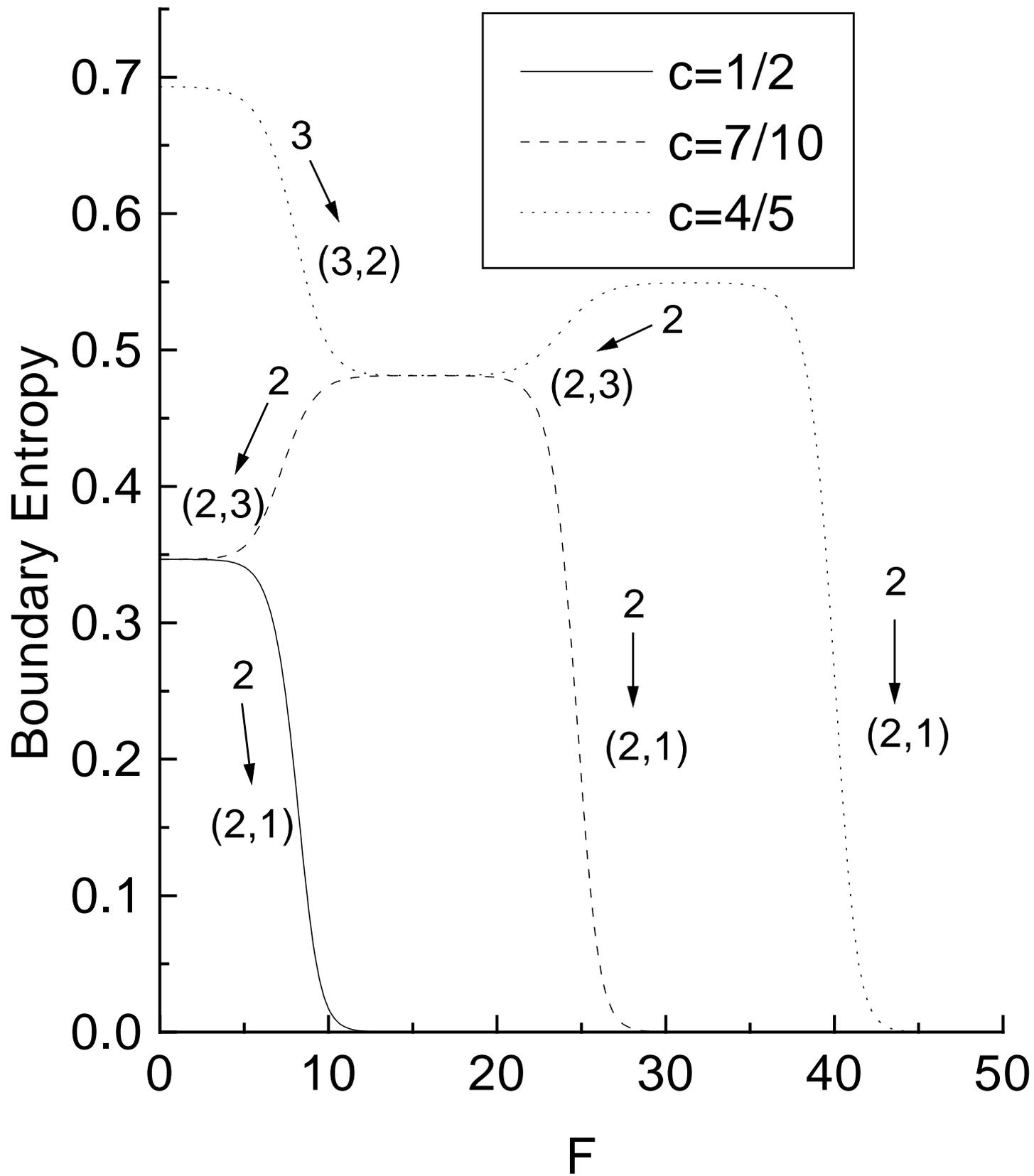
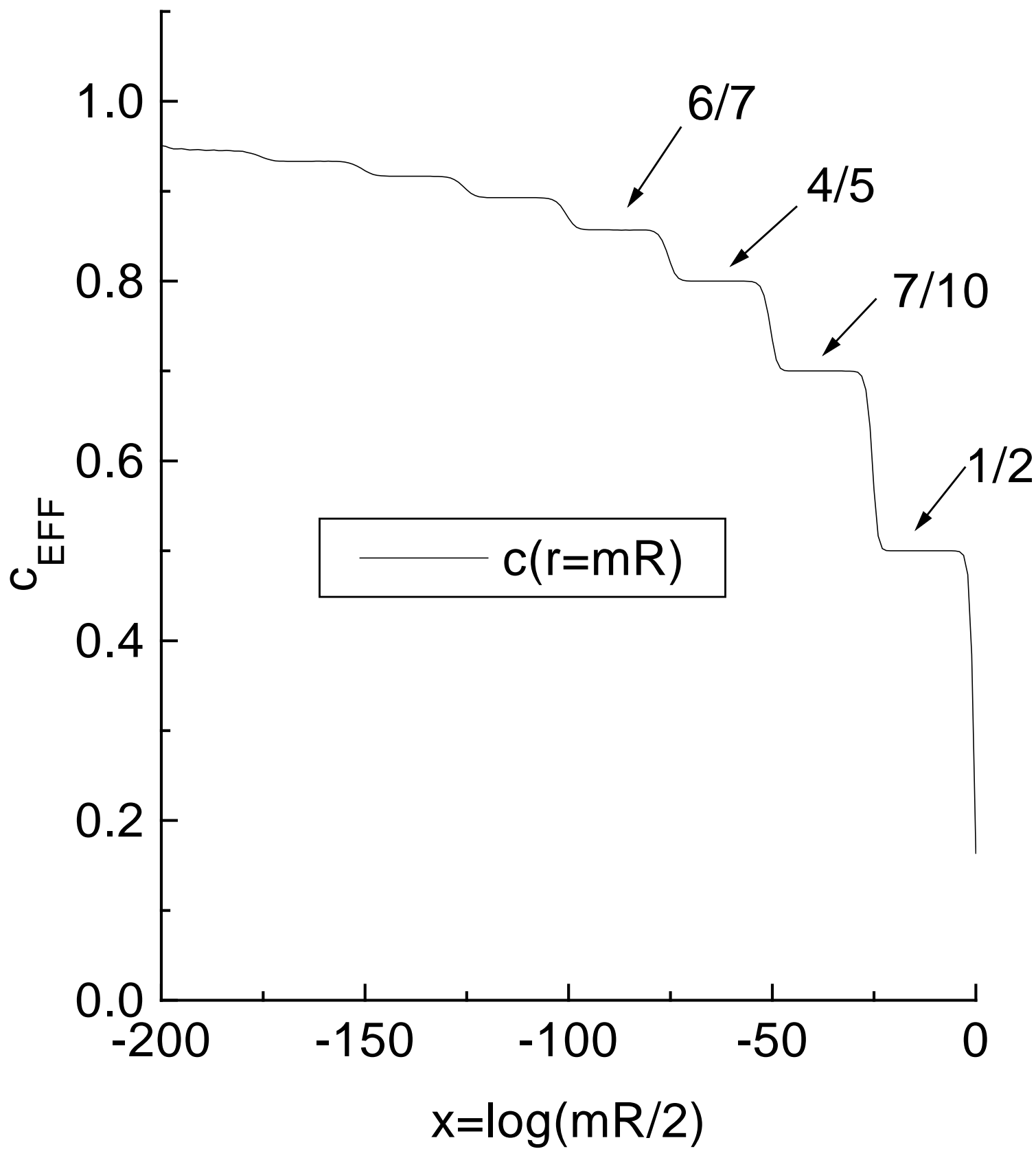
\bye